\documentclass[twocolumn]{aastex631}

\graphicspath{{./}{figures/}}

\usepackage{natbib}
\usepackage{graphicx}
\usepackage{amssymb}
\usepackage{epstopdf}
\usepackage{subfigure}
\usepackage{graphicx}
\usepackage{soul}

\usepackage{natbib}
\usepackage{graphicx}
\usepackage{amssymb}
\usepackage{epstopdf}

\newcommand{\bp}{{\mbox{\boldmath$p$}}}

\newcommand{\bvv}{{\mbox{\boldmath$v$}}}

\newcommand{\bx}{{\mbox{\boldmath$x$}}}

\newcommand{\bB}{{\mbox{\boldmath$B$}}}

\newcommand{\bE}{{\mbox{\boldmath$E$}}}

\newcommand{\om}{{\mbox{$\omega$}}}

\def  \be   {\begin{equation}}
\def  \ee   {\end{equation}}
\def  \beq  {\begin{eqnarray}}
\def  \eeq  {\end{eqnarray}}

\usepackage{color}

\def  \be   {\begin{equation}}
\def  \ee   {\end{equation}}
\def  \beq  {\begin{eqnarray}}
\def  \eeq  {\end{eqnarray}}

\begin{document}
\title{Chaotic Behavior of Trapped Cosmic Rays}


\correspondingauthor{Vanessa L\'opez-Barquero}
\email{vlopezb@umd.edu}

\author{Vanessa L\'opez-Barquero}
\affiliation{Institute of Astronomy, University of Cambridge, Madingley Road, Cambridge CB3 OHA, UK}
\affiliation{Department of Astronomy, University of Maryland, College Park, MD 20742-2421, USA}
\author{Paolo Desiati}
\affiliation{Wisconsin IceCube Particle Astrophysics Center (WIPAC), University of Wisconsin, Madison, WI 53703, USA}

\begin{abstract}
Recent experimental results on the arrival direction of high-energy cosmic rays have motivated studies to understand their propagating environment. The observed anisotropy is shaped by interstellar and local magnetic fields. In coherent magnetic structures, such as the heliosphere, or due to magnetohydrodynamic turbulence, magnetic mirroring can temporarily trap particles, leading to chaotic behavior. In this work, we develop a new method to characterize cosmic rays' chaotic behavior in magnetic systems using finite-time Lyapunov exponents. This quantity determines the degree of chaos and adapts to transitory behavior. We study particle trajectories in an axial-symmetric magnetic bottle to highlight mirroring effects. By introducing time-dependent magnetic perturbations, we study how temporal variations affect chaotic behavior. We tailor our model to the heliosphere; however, it can represent diverse magnetic configurations exhibiting mirroring phenomena. Our results have three key implications. (1)Theoretical: We find a correlation between the finite-time Lyapunov exponent and the particle escape time from the system, which follows a power law that persists even under additional perturbations. This power law may reveal intrinsic system characteristics, offering insight into propagation dynamics beyond simple diffusion. 
(2)Simulation: Chaotic effects play a role in cosmic ray simulations and can influence the resulting anisotropy maps. (3)Observational: Arrival maps display areas where the chaotic properties vary significantly; these changes can be the basis for time variability in the anisotropy maps. This work lays the framework for studying the effects of magnetic mirroring of cosmic rays within the heliosphere and the role of temporal variability in the observed anisotropy.

\end{abstract}

\section{Introduction}
\label{sec:intro}

The origin of the cosmic ray anisotropy observed over a wide range of energies~\citep{Abeysekara_2019,aartsen_2013chaos,2019ICRC...36..263G,2016ApJ...826..220A, 2018ApJ...865...57A,2018ApJ...861...93B,2017ApJ...836..153A} is still largely unknown. However, it is likely caused by a combination of factors. These factors include the spatial distribution of sources of cosmic rays in the Galaxy and the complex geometry and properties of the magnetic fields through which particles propagate. The processes shaping the distribution of cosmic rays are interconnected. Therefore, it is not trivial to unfold them~\citep{blasi_2012,DiSciascio_2015,Ahlers_2017,Deligny_2019,Gabici_2019,BeckerTjus_2020, evoli_2021}.

It has been speculated that the observed cosmic ray anisotropy in the 1-10 TV rigidity range may be explained in the context of homogeneous and uniform diffusion in the interstellar medium (ISM)~\citep{erlykin_2006,blasi_2012,ptuskin_2012,pohl_2013,sveshnikova_2013,savchenko_2015}. Nearby and recent sources are more likely to shape the cosmic rays' arrival direction distribution on Earth. On the other hand, the nonuniform pitch-angle distribution of the cosmic rays~\citep{effenberger_2012,mertsch_2015,giacinti_sigl_2012,tharakkal_2022} in magnetohydrodynamic (MHD) turbulence~\citep{barquero_2016} and the heterogeneous nature of the ISM affect the diffusion significantly in time and space. Therefore, the standard diffusion scenario cannot explain the complex angular structure of the observed anisotropy. Besides, nondiffusive stochastic scattering processes within the mean free path are likely to play an important role~\citep{giacinti_sigl_2012,ahlers_2014,ahlers_mertsch_2015,Battaner_2015, barquero_2016, Harding_2016, Kuhlen_2022, 2022arXiv220802261B}. \footnote{There are also more exotic scenarios to explain the small-scale anisotropy. In~\cite{Kotera_2013}, the authors propose that this anisotropy arises from the production of strangelets that could leave an imprint in the CR patterns. In~\cite{Harding_2013}, the origin will be from dark matter sub-halos.} The presence of coherent magnetic structures, such as superbubbles, magnetized clouds, or the heliosphere, can also cause a significant redistribution of the particle arrival directions.

Magnetic mirrors are present in a vast variety of astrophysical environments over a wide range of scales. Besides the heliosphere, coherent magnetic structures such as planetary magnetospheres, the Local Bubble, superbubbles, and likely galactic halos have a strong influence in trapping and redistributing cosmic rays. 
Spatial magnetic field intermittency, which plays a role in the formation of coherent structures~\citep{Matthaeus_2015,shukurov_seta_2017} and is involved in the transport and acceleration of charged particles, is consequently an important candidate to study when dealing with magnetic-bottle structures~\citep{Bell_2013}. Cosmic ray trapping in localized magnetic cells or mirrors may significantly contribute to the energy dependency of the diffusion coefficient~\citep{Hopkins_2021}. In particular, compressible modes in MHD turbulence generate the conditions for trapping cosmic ray particles, which leads to smaller and weaker energy dependency of diffusion parallel to the magnetic field lines~\citep{2020ApJ...894...63X}.

To study the fundamental processes occurring when particles are trapped by magnetic mirrors, we employ an idealized toy magnetic field system represented by an axially symmetric magnetic bottle (see Section~\ref{sec:mbottle}). Although this is an idealized system, it is known to cause complex particle trajectory topologies, and it serves the purpose of studying their properties. Particles may be permanently trapped within the magnetic bottle as long as their gyration frequency around magnetic field lines is sufficiently higher than the bouncing frequency between the mirror points. In such conditions, the magnetic field acting on particles does not significantly change within each gyration period. In other words, the motion is ``adiabatic." As soon as magnetic variations over each gyro-period start to become significant, the adiabatic limit breaks down and the particles' motion becomes increasingly complex. Trajectories may develop chaotic behavior, meaning that their deterministic geometry strongly depends on the initial conditions. 
All trajectories are deterministic and can be exactly determined as long as all aspects of the magnetic system as well as the particles' coordinates are known with infinite accuracy.
Even the slightest amount of inaccuracy makes any trajectory prediction impossible. Chaotic trajectories with similar initial conditions diverge from each other to very different trajectories. The rate of divergence depends on the actual initial conditions and the magnetic system, which determines the dynamic conditions to which trajectories are subjected. The degree to which similar trajectories diverge from each other can be assessed using the Lyapunov exponents. The variability of such exponents in the particles' phase space highlights the global properties of how a chaotic system is structured, 
and it may provide hints toward understanding how cosmic ray particles' arrival direction distribution on Earth is influenced by magnetic structures such as the heliosphere.

From a dimensional standpoint, cosmic rays with rigidity of $\sim$10 TV have a gyroradius of about $R_L\sim$ 500-800\,AU in a 3-5\,$\mu$G magnetic field, which is comparable to the transverse size of the heliosphere~\citep{pogorelov_2013}. In fact, while low-rigidity cosmic rays are influenced by the inner heliospheric structure, 10 TV scale particles are shaped by the boundary region with the ISM~\citep{desiati_lazarian_2013,schwadron_2014,zhang_2014,barquero_2017}.

Therefore, it is evident that to determine the cosmic rays' distribution in the interstellar medium, it is necessary to account for the heliospheric influence~\citep{zhang_2014,barquero_2019,snowmass2021}. Currently, we seem to know more about the inner heliosphere, while little is understood about the interface between the solar wind and the local ISM. Various questions arise: How wide and long is the heliosphere? Are the flanks characterized by magnetic instabilities? Does turbulence play a relevant role? Therefore, a careful analysis of experimental observations, along with the most up-to-date heliosphere models, may help account for the heliospheric effects on arriving cosmic rays. The recent full-sky combined observation of the 10-TeV cosmic ray anisotropy by the HAWC gamma-ray and the IceCube neutrino observatories~\citep{Abeysekara_2019} provides the first view of TeV cosmic ray anisotropy with minimal experimental bias~\citep{barquero_2019}.

In~\cite{barquero_2017}, protons, helium, and iron nuclei trajectories between 1 TV and 10 TV were numerically integrated in a heliospheric magnetic field model by~\cite{pogorelov_2013}. There is no turbulence or stochastic magnetic field in the model. However, despite that, the initial uniform arrival direction distribution from the local interstellar magnetic field is broken down into medium and small angular scales by the effects of the heliospheric magnetic bubble. The corresponding angular power spectrum is not different from that generated by scattering processes off compressible MHD turbulence~\citep{barquero_2016}. It turns out that cosmic ray particles with rigidities of 1–10 TV may be temporarily trapped in the magnetic mirror configuration formed by the interstellar magnetic field lines draping around the heliosphere flanks.
%

In Section~\ref{sec:mbottle}, we present the physical contexts where the studies of particles in a magnetic bottle are laid down. Section~\ref{sec:propChaos} describes how particle trajectories are numerically calculated.  
Section~\ref{sec:chaos} introduces the aspects about chaos theory that are relevant for this work, with Section~\ref{sec:lya} describing the Lyapunov exponents as an estimate of the degree of chaos in a system. Results are presented in Section~\ref{sec:results} and discussed in Section~\ref{sec:disc}. The connection with the observations is given in Section \ref{sec:observations}. The outlook is given in Section~\ref{sec:outlook}. The summary and conclusions are given in Section~\ref{sec:conclusions}.


\section{The Magnetic Bottle Field}
\label{sec:mbottle}

%
\begin{figure*}[t]
	\centering
	\includegraphics[width=\linewidth]{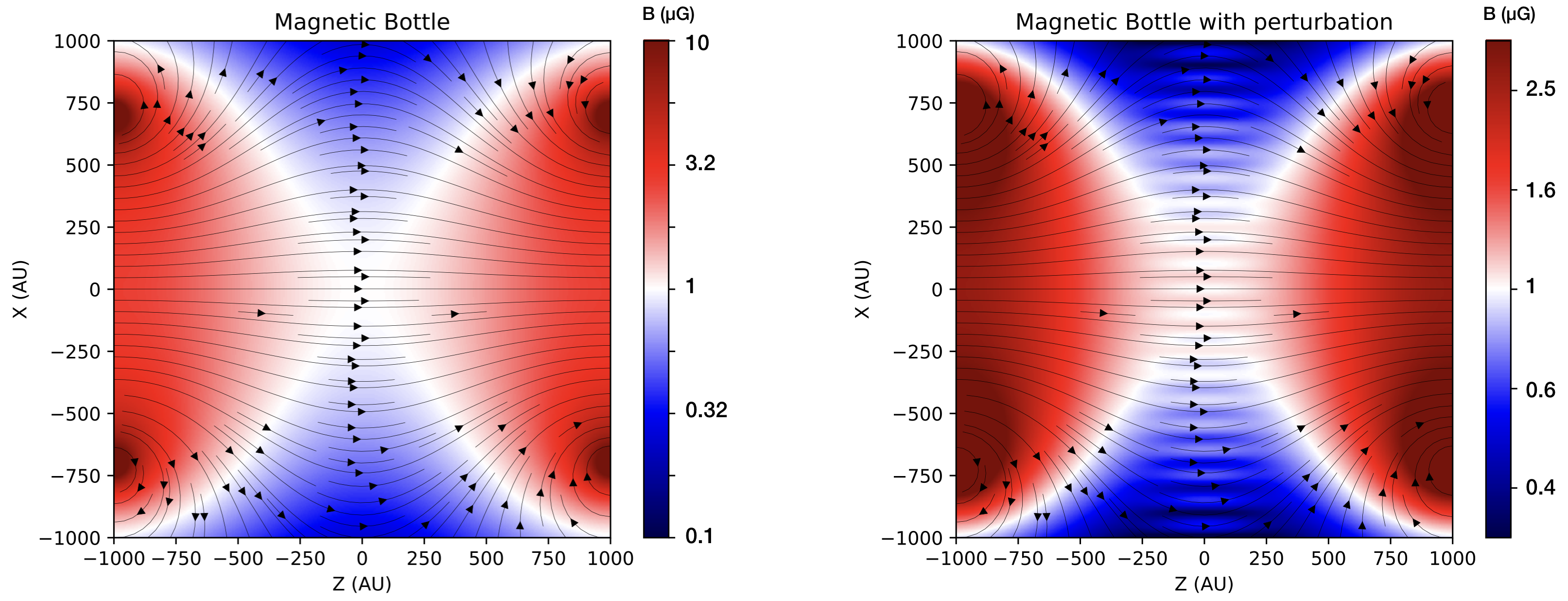}
	\caption{The magnetic bottle field geometry used as toy model to study the behavior of particles trapped by the interstellar magnetic field draping around the heliosphere. On the left, the static magnetic field, and on the right, with the additional perturbation imitating the effects of solar cycles on the heliospheric magnetic field along its tail.}
	\label{fig:mbottle}
\end{figure*}
An axial-symmetric magnetic bottle is used as a toy model to study how cosmic rays are trapped and under which conditions their trajectories' chaotic behavior arises and develops. 

This magnetic field is generated by two circular coils with electric currents running in the same direction. Although the purpose of using this toy model goes beyond the investigation of TeV cosmic rays in the heliosphere, as detailed in Section~\ref{sec:outlook}, we tailor our model to replicate the features of the heliospheric system. We assume that its spatial scale is comparable to the size of the heliosphere and adjust the magnetic field magnitudes to approximately those of the heliosphere. In this regard, we pick the distance between the coils as 2000 AU, which is the scale at which the local interstellar magnetic field lines drape around the heliosphere~\citep{pogorelov_2013}. The coils' radius and currents are selected so that the magnetic field is approximately 3 $\mu$G at the center of each coil (corresponding to the mirror points of the magnetic bottle) and the lowest possible at the center between the two coils. Such a condition is satisfied with the geometric parameters listed in Table~\ref{tab:geometry}.
\begin{table}
	\centering
	\caption{Parameters for the Magnetic Bottle}
	\label{tab:geometry}
	\begin{tabular}{ | p{3cm} |  p{3cm} | }
		\hline \hline
		Radius (R) & 700 AU \\ 
		\hline
		Current (I) & $ 4 \times 10^{10} $ A \\
		\hline
		Distance (D) & 2000 AU \\
		\hline
	\end{tabular} 
\end{table}
%
%
With these parameters, the magnetic field is about 2.7 $\mu$G at the center of the coils and about 1 $\mu$G at the point between the coils. A cross section representation of the resulting field is shown on the left in Figure~\ref{fig:mbottle}, where the magnetic field intensity is shown in color scale and the magnetic field line shows the shape of the magnetic bottle.
\begin{figure*}
	\centering
	\includegraphics[width=0.48\linewidth]{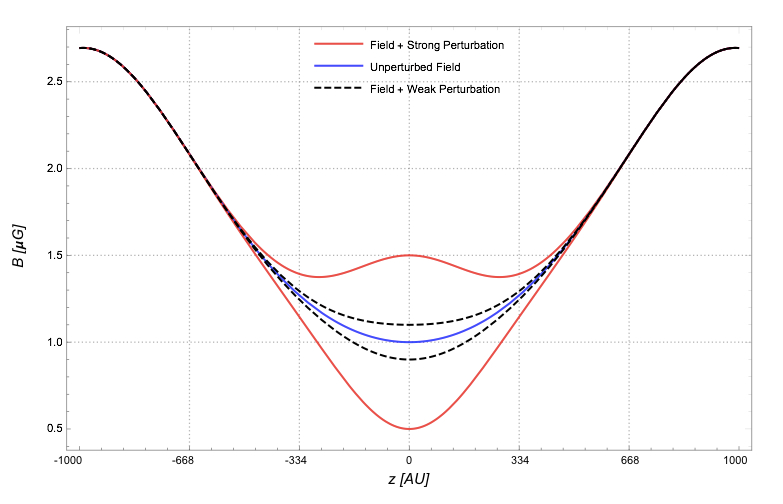}
	\includegraphics[width=0.5\linewidth]{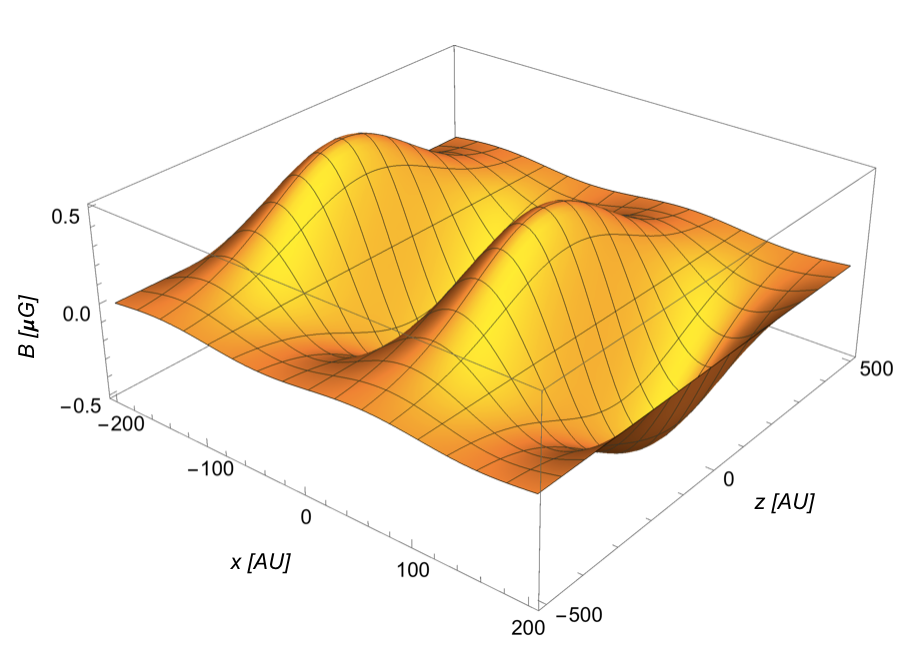}
	\caption{On the left, the field profile along the axis of the magnetic bottle with the {\it weak} and {\it strong} perturbations at their maximum amplitude. On the right, the 3D view of a snapshot of the magnetic perturbation.}
	\label{fig:magneticfieldstrongweakunperturbed}
\end{figure*}


Time-dependent magnetic perturbations are employed to investigate the impact of external effects on chaotic behavior. With the heliospheric system as inspiration, we introduce time modulations that mimic the effects of magnetic field reversals induced by the 11-year solar cycles. To do so in our toy model, we add a time-dependent component propagating transversely through the magnetic bottle (along the x-axis) with periodic modulations along the y-axis and a Gaussian dependency along the magnetic bottle axis (the z-axis) so that the largest perturbation is located at the center of the magnetic system. Such magnetic perturbation, shown on the right of Figure~\ref{fig:mbottle}, is represented by the function
\begin{equation}
B_{y} = \frac{\Delta B}{B}\,\sin(k_p x-\omega_p t)\,e^{-\frac{1}{2}\left(\frac{z}{\sigma_p}\right)^2},
\label{eq:pert}
\end{equation}
where $k_p = \frac{2\pi}{L_p}$ and $\omega = \frac{2\pi v_p}{L_p}$ with $L_p$ = 200 AU the spatial scale of the magnetic polarity regions, and $\sigma$ = 200 AU the width of the Gaussian modulation of the perturbation (see Figure~\ref{fig:magneticfieldstrongweakunperturbed}). 
The relative amplitude $\frac{\Delta B}{B}$ and velocity $v_p$ depend on the strength and type of magnetic perturbation, as shown in Table~\ref{tab:param}.
\begin{table}
	\centering
	\caption{Parameters for the Time-Perturbation}
	\label{tab:param}
	\begin{tabular}{ | p{2.0cm} |  p{1.5cm} | p{1.5cm} |p{1.5cm} | }
		\hline \hline
		 & Weak &Weak + E & Strong   \\ 
		\hline
		$\frac{\Delta B}{B}$ &  0.1 & 0.1 & 0.5\\ 
		\hline
		$v_p$ (AU/yr) & 2 & 2 & 20 \\
		\hline
	\end{tabular} 
\end{table}
%
The {\it weak} perturbation approximately represents the variability of solar wind properties along the heliosphere beyond the termination shock (see~\cite{Pogorelov_2009}). The parameters for the {\it strong} perturbation are chosen to amplify the effects of magnetic field time-modulations on the properties of particle trajectories.

When a magnetic field changes in time, an induced electric field $\bE = -\bvv\times \bB$ is produced. However, in plasmas, the induced electric fields are typically very small because the high conductivity makes it possible for electric charges to rearrange and screen electric fields over distances larger than the Debye length. The Debye length is the distance over which the screening caused by the collective charge rearrangement is effective, and shielding can occur only if the Debye length is much larger than the average distance of particles in the plasma. The heliospheric plasma has a wide variability of its properties, and it is difficult to pinpoint specific numbers that represent the global heliospheric behavior. In this work, we assume the extreme scenario where electric fields induced by the {\it weak} magnetic perturbations are not screened. With the parameters described in Table \ref{tab:param}, the magnitude of the force produced by the electric field compared to the one from the magnetic field is approximately three orders of magnitude smaller. Consequently, no significant effects from the electric field's presence are expected. The possibility for the occurrence of electric fields and its effects on the observed anisotropy is also studied in~\cite{drury_2013}.




\section{Calculating particle trajectories}
\label{sec:propChaos}


Particle trajectories are calculated by numerically integrating the equation of motion
\begin{equation}\label{momentum}
\left\{
\begin{array}{lcl}
\frac{d\bp}{dt}=q\left(\bE+\frac{\bvv\times\bB}{c}\right) \\
\frac{d\bx}{dt} = \bvv
\end{array}
\right.,
\end{equation}
describing the force exerted by an electric field $\bE$ and magnetic field $\bB$ on a particle with velocity $\bvv$ and momentum $\bp$. 
As in~\cite{desiati_zweibel_2014}, a dimensionless version of Eq.~(\ref{momentum}) is used in this work, where we introduce a magnetic field scale $B_0$,
\begin{equation}\label{dimlessmomentum}
\left\{
\begin{array}{lcl}
\frac{d\hat\bp}{ds}  =  \hat\bE+\frac{\hat\bp}{\gamma}\times\hat\bB \\
\frac{d\hat \bx}{ds} =  \frac{\hat\bp}{\gamma}.
\end{array}
\right.,
\end{equation}
where $\hat\bB\equiv\bB/B_0$ and $\hat\bE\equiv\bE/c B_0$ are the normalized magnetic and electric fields, respectively, and $\om_0\equiv eB_0/m_p$ is the proton gyrofrequency scale, which defines the dimensionless time $\hat{t}\equiv\om_0t$. The gyroradius scale $r_0\equiv c/\om_0$ defines the dimensionless spatial coordinates $\hat\bx\equiv\bx/r_0$, while the dimensionless momentum is defined as $\hat\bp\equiv\bp/mc$.
The particle velocity $\bvv$ is related to $\hat\bp$ by $\bvv = \hat\bp/\gamma$ and its Lorentz factor $\gamma = \sqrt{1+\hat p^2}$. In these units, the dimensionless particle gyroradius is $\hat r_g = \hat p_{\perp}$, and the dimensionless gyrofrequency is $\hat \omega_g = 1/\gamma$. Normalized variables are written with hats. 

Eqs. (\ref{dimlessmomentum}) are numerically solved using the fourth order Runge-Kutta integration method, with an adaptive time-step-size algorithm that keeps relative truncation errors within a tolerance level of $\epsilon = 10^{-10}$ (see~\cite{desiati_zweibel_2014} for more discussion on numerical accuracy). The maximum integration time used in this work was set to $\hat{t}_{max} = 10^8$ in code units (corresponding to about 10$^{10}$ seconds, or 330 years). Under these conditions, the accuracy of the numerical integration is sufficient and does not affect the results.
The magnetic field configurations described in Section~\ref{sec:mbottle} are used to calculate 1 TeV antiproton trajectories propagating back in time from their final location, at coordinates ($\hat{x}_0$, $\hat{y}_0$, $\hat{z}_0$) = (100, 100, 500) in code units, away from the symmetry point of the magnetic system geometry. Integration stops either when integration time reaches the maximum value of $\hat{t}_{max} = 10^8$ or when the trajectories cross a sphere centered on ($\hat{x}$, $\hat{y}$, $\hat{z}$) = (0, 0, 0) with radius $r_{max}$ = 12500 in code units, corresponding to 2500 AU. Four sets of trajectories were calculated: one with the static magnetic bottle configuration shown on the left of Figure~\ref{fig:mbottle}, one with the addition of the {\it weak} magnetic perturbation of Eq.~\ref{eq:pert}, one with the {\it strong} magnetic perturbation, and the last using the weak magnetic field perturbation and the induced electric field $\bE = -\bvv_{p}\,\times \bB$. For each set, a total of 768 antiproton trajectories were integrated, with momentum vector direction corresponding to each pixel in a HealPix grid~\citep{gorski_2005} with $nside = 8$.

To study the onset of chaotic behavior, i.e., how trajectories with infinitesimally close initial conditions diverge from each other, we produce, for each of the 768 {\it reference} trajectories of the four sets, ten additional sets of trajectories with the same initial momentum and with initial position randomly distributed around ($\hat{x}_0$, $\hat{y}_0$, $\hat{z}_0$) = (100, 100, 500) on a sphere of radius $\hat{r}_0$ = 0.01.




\section{Chaotic Trajectories}
\label{sec:chaos}

All physical systems that are conservative can be described as Hamiltonian systems, where the total energy and phase-space volume are conserved. One of the properties of Hamiltonian systems is that their state is governed by deterministic laws. These systems can be highly sensitive to initial conditions, which is what characterizes chaotic systems. Even the smallest differences in the initial conditions, whether they effectively originate from measurement uncertainties or from rounding errors of numerical calculations, may lead to vastly different trajectories. The limited knowledge of the properties of a physical system, in addition to experimental or numerical resolution and accuracy, makes long-term prediction of its state evolution generally impossible, despite its deterministic nature. In a chaotic system, for an arbitrarily small solid angle in the sky, the origin of the particles coming from it can be highly uncertain and unpredictable.
%
%
In the classical approximation, chaos can explain the origin and mechanisms of apparently stochastic processes, and this deterministic randomness can occur even in a very limited number of degrees of freedom.
%



A known chaotic system is the axis-symmetric magnetic bottle~\citep{Chirikov_1987,Ambashta_1988}. Particles trapped in a magnetic bottle are characterized by their gyration frequency around the magnetic fieldlines and their bouncing frequency between the mirror points. As long as gyration frequency is sufficiently higher than bouncing frequency, the magnetic force on the particles changes very slowly within each gyration. In this condition, the magnetic moment $\mu = \frac{1}{2}\frac{p_{\perp}^2}{B}$ is an approximate constant of motion. In the limit of perfect conservation of the magnetic moment (also known as first adiabatic invariant), particles are indefinitely trapped inside the magnetic bottle and bounce back and forth between the mirror points. In reality, since the the adiabatic invariant is never exact in a realistic magnetic field system, after a sufficient time, particles eventually escape from the system. When the gyration and bouncing frequencies, which can slowly drift from their initial values, assume comparable values, the changes in the magnetic field during one gyration may no longer be negligible and the adiabatic condition may be violated~\citep{Ambashta_1988}. In~\cite{Chirikov_1987}, the case of chaotic adiabaticity is discussed for a set of initial conditions. Even though the idea of adiabatic chaos may be counter-intuitive, it is reasonable to think that a system that varies slowly can preserve its adiabaticity while having divergent trajectories~\citep{Jarzynski_1994}.

Particles with very close but separate trajectories in phase-space experience different magnetic forces that eventually pull them apart. Under certain conditions, the rate of separation of similar trajectories is significant so that they continue to develop with very different topologies before escaping the system. When this happens, trajectories manifest a typical chaotic behavior, which determines how long it takes before particles escape. 


\subsection{Lyapunov Exponents}
\label{sec:lya}

One way to characterize chaotic trajectories is through the Lyapunov exponents (LE)~\citep{1994PhRvL..73.1927D,McCue_2011,2005PhLA..335..394S,1985PhyD...16..285W}. As particles with an initial separation propagate, they will start to get farther apart, closer together, or remain at a constant separation; therefore, the LE will quantify the rate of divergence or convergence of the trajectories. If $ \delta Z_{0} $ is the initial separation and $ \delta Z (t)$ is the separation at time $t$, these two quantities can be related by the expression

\begin{equation}\label{lyaexp}
\left | \delta Z (t)  \right | \approx e^{\lambda t} \left | \delta Z_{0}  \right | ,
\end{equation}
where $\lambda$ is the Lyapunov exponent. Accordingly, a negative LE indicates convergence and a positive one, divergent trajectories and possibly chaos.



The number of Lyapunov exponents in the spectrum will depend on the dimensionality of the phase space. The largest exponent is referred to as the maximal Lyapunov exponent (MLE). This exponent will eventually dominate over the others due to exponential growth. Typically, the MLE is used to describe the trajectories since it is relatively simple to calculate from a time series and information can be obtained readily from it. The MLE can be expressed as

\begin{equation}\label{maxLyap}
\lambda = \lim_{t\to\infty}  \lim_{\delta Z_{0} \to 0} \frac{1}{t}   \ln \frac{\left | \delta Z (t)  \right |}{\left | \delta Z_{0}  \right | },
\end{equation}
where effectively the initial separation is made as small as possible and an asymptotic behavior is sought taking the limit of $t$ to infinity. 

However, problems arise when we look for such asymptotic behavior, since a trajectory may never achieve it, e.g., if the particle moves from one environment to another in a short time or it gets affected by different first-order mechanisms on its way. One way to alleviate this problem is to use the finite-time Lyapunov exponent (FTLE). Through the FTLE, a finite-time interval can be used to calculate the divergence in the trajectories without the necessity of an infinite limit. 

The FTLE expression is given by

\begin{equation}\label{FTLE}
\lambda (t,\Delta t)=   \frac{1}{\Delta t}   \ln \left [ \frac{d(t+\Delta t)}{d(t) } \right ],
\end{equation}
where $\Delta t$ is the time interval for the calculation. The value for  $\Delta t$ is chosen depending on the intrinsic characteristics of the system and the particles traveling through it. Therefore, it is flexible and can be adapted to different scenarios. 

In this work, we are dealing with a bounded system; therefore, it is imperative to use a quantity that will quantify chaos under such conditions. Since an asymptotic behavior is not achieved for particles that remain in the system for a period of time before escaping, the FTLE can adjust and describe their behavior while bounded in the system, given that an appropriate $\Delta t$  is chosen. 



\section{Methodology}
\label{sec:method}

\begin{figure*}[t]
	\centering
	\includegraphics[width=0.8\linewidth]{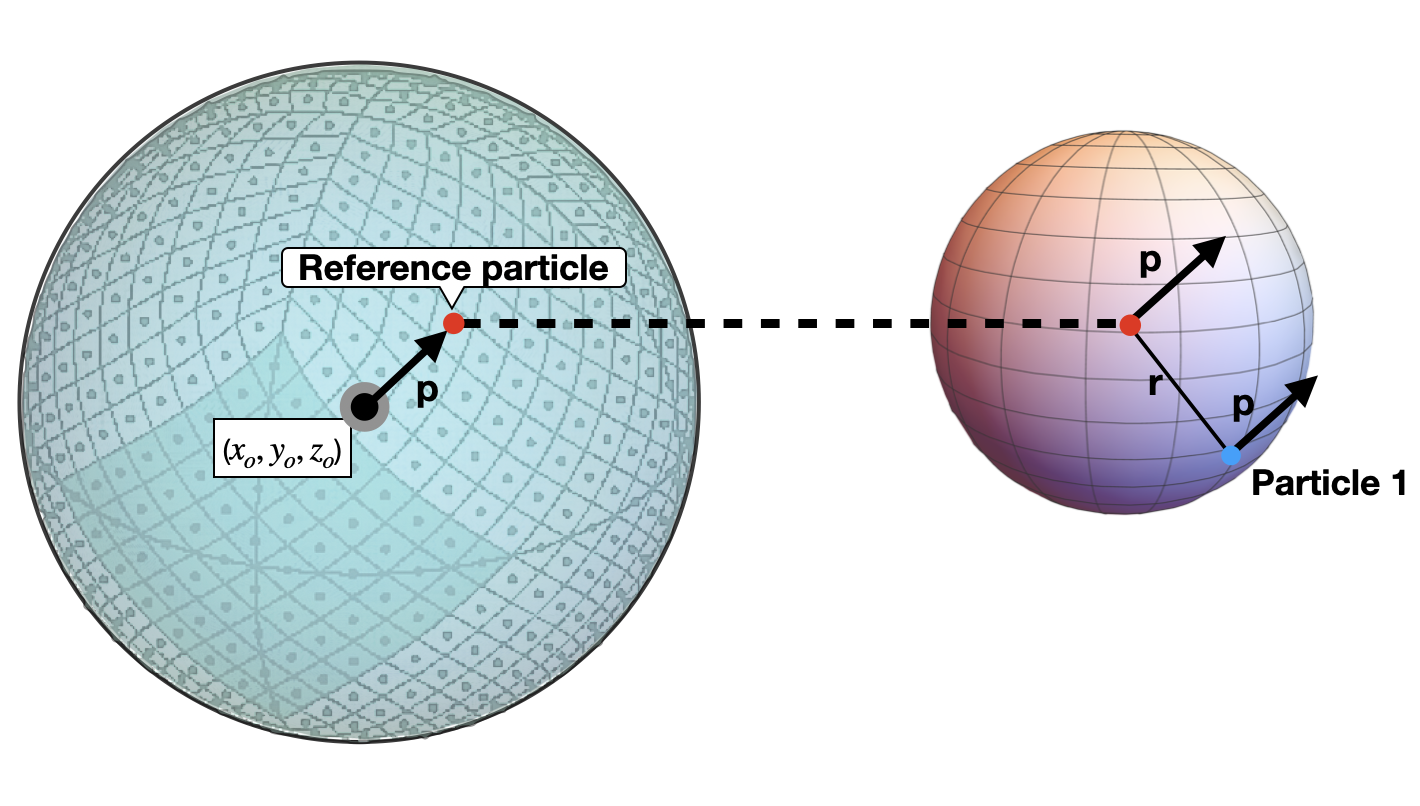}
	\caption{Method for the injection of particles. We introduce our reference particles starting at position ($x_o$,$y_o$,$z_o$)=(100, 100, 500) and with initial momentum in the direction of the 768 pixels in the map, which correspond to each pixel in the HealPix grid with $nside = 8$ (Figure on the left modified from~\cite{Gorski_99}). This figure shows an example of a reference particle denoted by a red dot, and the light blue sphere represents the possible momenta directions. For each reference particle (red dot), we have a set of 10 particles injected randomly on the surface of a sphere of radius $r$ = 0.01. On the right, we have an example of a reference particle with one of the particles in its family (blue dot). These two particles are separated a distance of 0.01 and have identical momenta. The main idea behind this process is to inject particles with almost identical initial conditions to track how chaotic these trajectories are.}
	\label{fig:method}
\end{figure*}

\begin{figure*}
	\centering
	\includegraphics[width=1\linewidth]{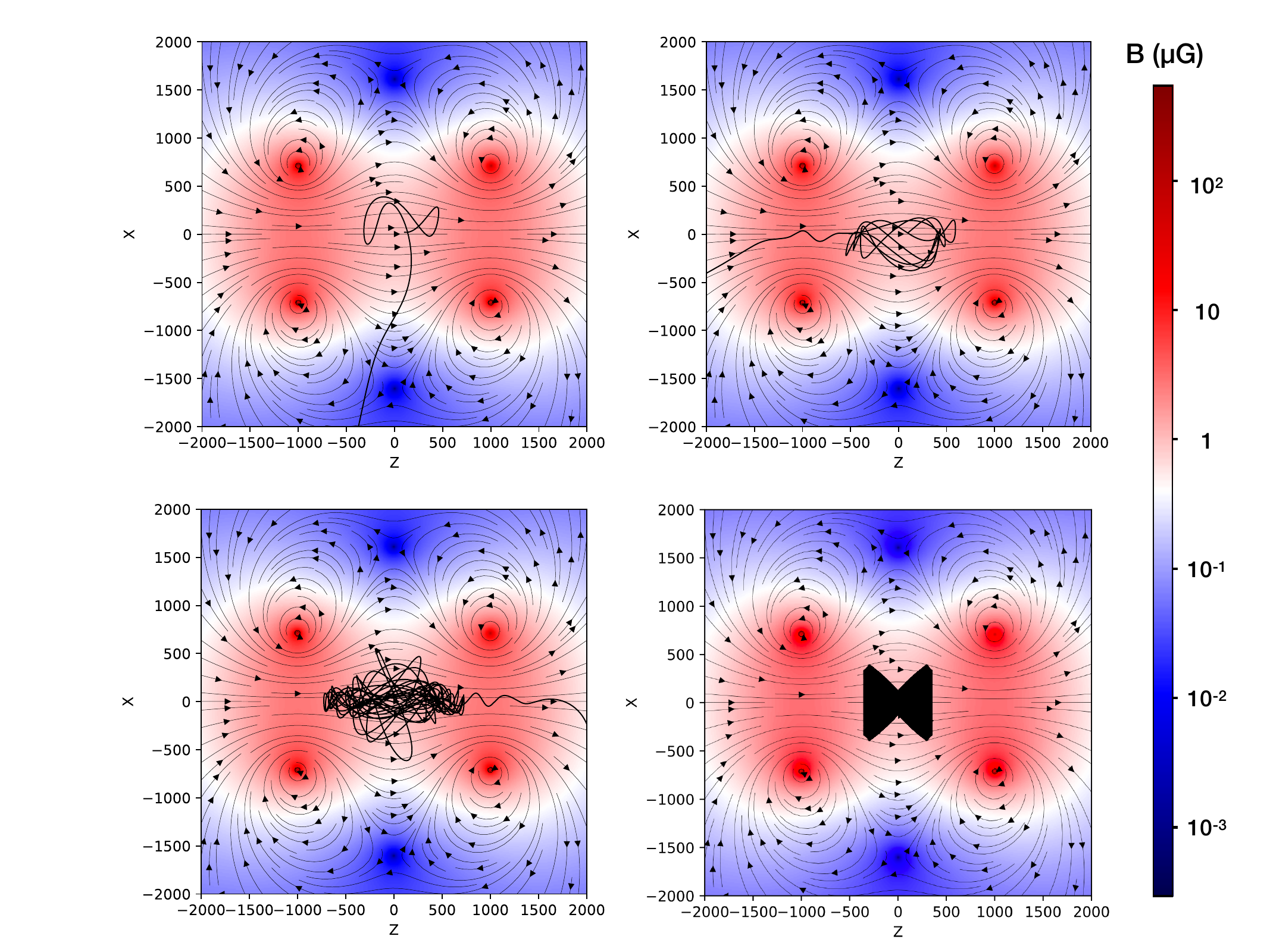}
	\caption{Trajectories in the unperturbed system. \textit{Top Left}: 
	Transient particle with a final time of 33000.  \textit{Top Right}: 
	Intermediate particle with a final time of 75402.  \textit{Bottom Left}: 
	Irregular particle in the power-law behavior section with a final time of 295366. \textit{Bottom Right}: 
	Trapped particle with the maximum integration time. The transient and trapped particles do not display chaotic behavior, whereas the intermediate and power-law behavior particles are chaotic.}
	\label{fig:figure-trajectories}
\end{figure*}

We introduce our reference particles starting at the point ($x_o$,$y_o$,$z_o$)=(100, 100, 500) and with initial momentum in the direction of the 768 pixels in the map, which correspond to each pixel in the HealPix grid~\citep{gorski_2005} with $nside = 8 $.
For each reference particle, we have a set of 10 particles that are injected randomly on a sphere of radius $\hat{r}_0$ = 0.01 and with the same momentum (magnitude and direction) as their reference particle (see Figure \ref{fig:method}). 
The final time for the family of particles per reference particle is defined as the shortest final time for a specific particle.
At each time step, the distance in phase space is calculated between each particle and the reference, given by the expression

\begin{equation}\label{distance}
d^{2}_j(t) = \sum_{i=1}^{3} (x_{i}^{ref}-x_i)^2 +(p^{ref}_i-p_i)^2 .
\end{equation}
With all the distances calculated, we proceed to the calculation of the finite-time Lyapunov exponent, given by the expression
\begin{equation}\label{calc_lya}
\lambda_{j} =   \frac{1}{\Delta T}   \ln \left [ \frac{d(t_{j}+\Delta T)}{d(t_{j}) } \right ] .
\end{equation}

\begin{figure*}
	\centering
	\includegraphics[width=1\linewidth]{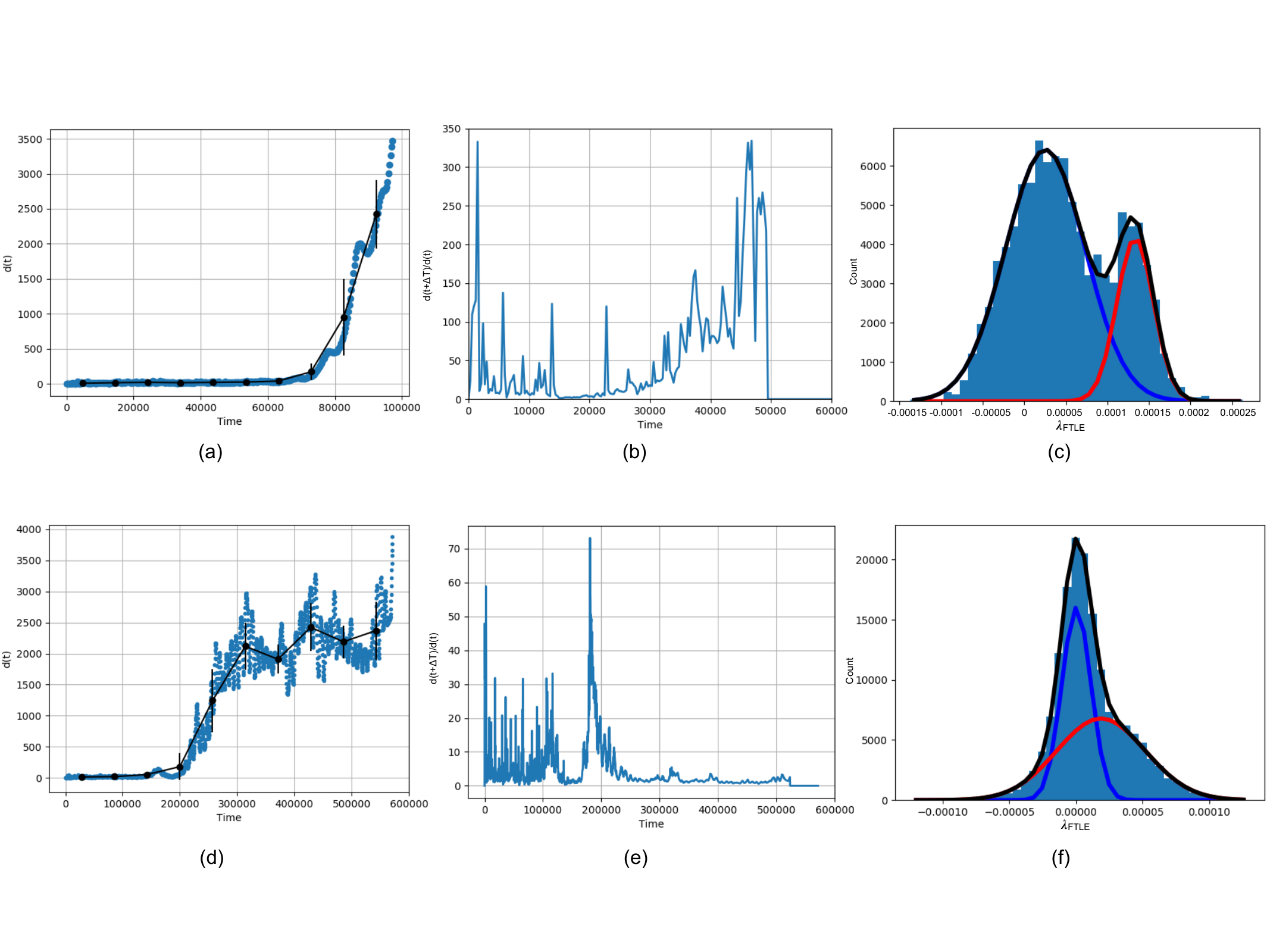}
	\caption{ Comparison between the behavior of two particles with different escape times.
    \textit{Top Panels}: These correspond to a particle with an escape time $t_{esc}= 97251$.
    \textit{Bottom Panels}: Particle with an escape time $t_{esc}= 5.7 \times 10^5$. 
    Panels (a) and (d) show distance in phase space vs. time; panels (b) and (e) show distance in phase space at time $t + \Delta T$ over the distance at time $t$ as a function of time $t$; panels (c) and (f) display a histogram of the finite-time Lyapunov exponent $\lambda_{\tiny{FTLE}}$ and the corresponding fits denoted with black, red, and blue lines; see \ref{sec:method} for details on the Gaussian fits. Note that for the shorter trajectories (a), they stay with almost no separation for a short time and then diverge rapidly and leave the system right away. On the contrary, longer trajectories (d) take longer to start diverging and when they do, the process takes a longer time with intermediate periods of slower divergence before they are able to escape. 
        }
	\label{fig:sepratiohisto206216}
\end{figure*}

The value for $\Delta T$ should be chosen depending on the characteristics of the system; for example, in this case, the bouncing time between mirrors gives us a point of reference for the value of $\Delta T$.  
Also, $\Delta T$ should capture the specific features of the divergence, as shown in panels (a) and (d) in Figure \ref{fig:sepratiohisto206216}. Here, the average value taken for $\Delta T$ is 38100 in normalized units. 

Once the FTLE has been calculated for each pair of particles at each time step $t_{j}$, we proceed to calculate the average for the family of particles at each time step. Therefore, we obtain a $\bar{\lambda}$ for each time step: 

\begin{equation}\label{calc_lya_ave}
\bar{\lambda}^F_j = \frac{1}{n} \sum_{i=1}^{n} \lambda_{ij} ,
\end{equation}
with n as the number of particles in the family for each reference particle; in our case, n=10. 

Then, a histogram is generated with all the obtained values of $\lambda$. Panels (c) and (f) in Figure \ref{fig:sepratiohisto206216} show in light blue an example of an obtained distribution. Given that a value of $\lambda_{FTLE}$ equal to zero means no divergence and that a positive value indicates divergence, we then proceed to fit two Gaussians for each distribution (see~\cite{2005PhLA..335..394S}). As we can see from our two examples in Figure \ref{fig:sepratiohisto206216}, there is a peak in the distribution around zero and another peak at a higher value. 
Therefore, we fit these two scenarios with the peaks of our two Gaussians. Since the second peak represents the actual divergence of the trajectories, we take this value to represent the value assigned to $\lambda_{FTLE}$ for each specific family of particles.



\section{Results}
\label{sec:results}

This section shows the results obtained with the numerical calculation and methodology described in Sections \ref{sec:propChaos} and \ref{sec:method}, respectively. Based on those, it is found that there is a correlation between the finite-time Lyapunov exponent (FTLE), i.e., the chaotic behavior of the set of particles, and the escape time from the system (see Fig. \ref{fig:finitelyapunovvsescapetimeunperturbedchi2500}). This correlation follows a specific power law that persists even if perturbations are introduced in the system (see Figs. \ref{fig:finitelyapunovvsescapetimefourcaseschi2500} and \ref{fig:finitelyapunovvsescapetimemigration}). If the FTLEs and escape times are plotted in arrival distribution maps, we observe that regions with different chaotic behavior emerge as well as gradients that appear between them (see Fig. \ref{fig:figure-skymaps}).

\subsection{Classification of Particles}
\label{ssec:class}

Given our analysis for each set of particles, we can classify them based on the behavior of the finite-time Lyapunov exponent and its relation to the escape time, i.e., the time that the set of particles spent in the system. 

In these systems, we can identify five different regimes based on the particles' behavior. The particles with the shortest final times are transient (see Fig. \ref{fig:figure-trajectories}). These are particles that have a final time lower than 50000 (in code/normalized units) and their initial position is around the equator in the maps in Figure \ref{fig:figure-skymaps}. They do not spend much time in the magnetic system and these trajectories pass through regions with no strong variations of magnetic field intensity. Since they escape the system very quickly, they do not have time to develop any chaos while in the system. 


Particles with final times between 50000 to $10^5$ are in an intermediate state (see Fig. \ref{fig:figure-trajectories}). Particles in this intermediate state diverge so quickly that the system can not contain them, and therefore they cannot reach a steady state for their chaotic behavior. These particles tend to have the highest values of $\lambda_{\tiny{FTLE}}$, especially in the perturbed cases. 


The great majority of particles have final times between $10^5$ and $10^8$ (see the histogram in Figure \ref{fig:histoselectedfinaltimesloglmandtrfunpert}). These trajectories are chaotic and consequently sensitive to the initial conditions. Their behavior follows a power law that correlates the escape time and the Lyapunov exponent. 


The particles that fit this power law behavior can be subdivided in two categories depending on their chaotic attributes, irregular and regular. Irregular particles have final times between $10^5$ and $10^{5.5}$. The divergence of these particles is sudden, and they do not experience a steady state as the regular particles do (see panel (a) in Figure \ref{fig:sepratiohisto206216}). 

Regular particles, with final times between $10^{5.5}$ and $10^8$, start to diverge at a slower pace compared to the irregular particles. Later, after a period of divergence, they achieve a steady state. They spend most of the time in this steady state, and then they leave the system  (panel (d) in Figure \ref{fig:sepratiohisto206216}). These trajectories are long trajectories at the margin of the stability region in phase space. 

The final category is trapped particles. These particles are only present in the unperturbed system. Their final time is our maximum value of $10^8$. These trajectories occupy the stability region, which is the region in phase space where trajectories remain trapped within the integration time (here $10^8$).  These trajectories, despite being very long, are not sensitive to initial conditions, and they are stable. 
They will be trapped in the magnetic mirror if there is no time-dependent perturbation. With time-dependent magnetic perturbations, these trajectories lose their adiabatic properties and escape after a relatively long bounded period. This change depends on the strength of the perturbation, as we will see in the next subsection. 




\subsection{Finite-Time Lyapunov Exponents vs. Escape Times}
\label{ssec:ftle}

\begin{figure*}
	\centering
	\includegraphics[width=0.75\linewidth]{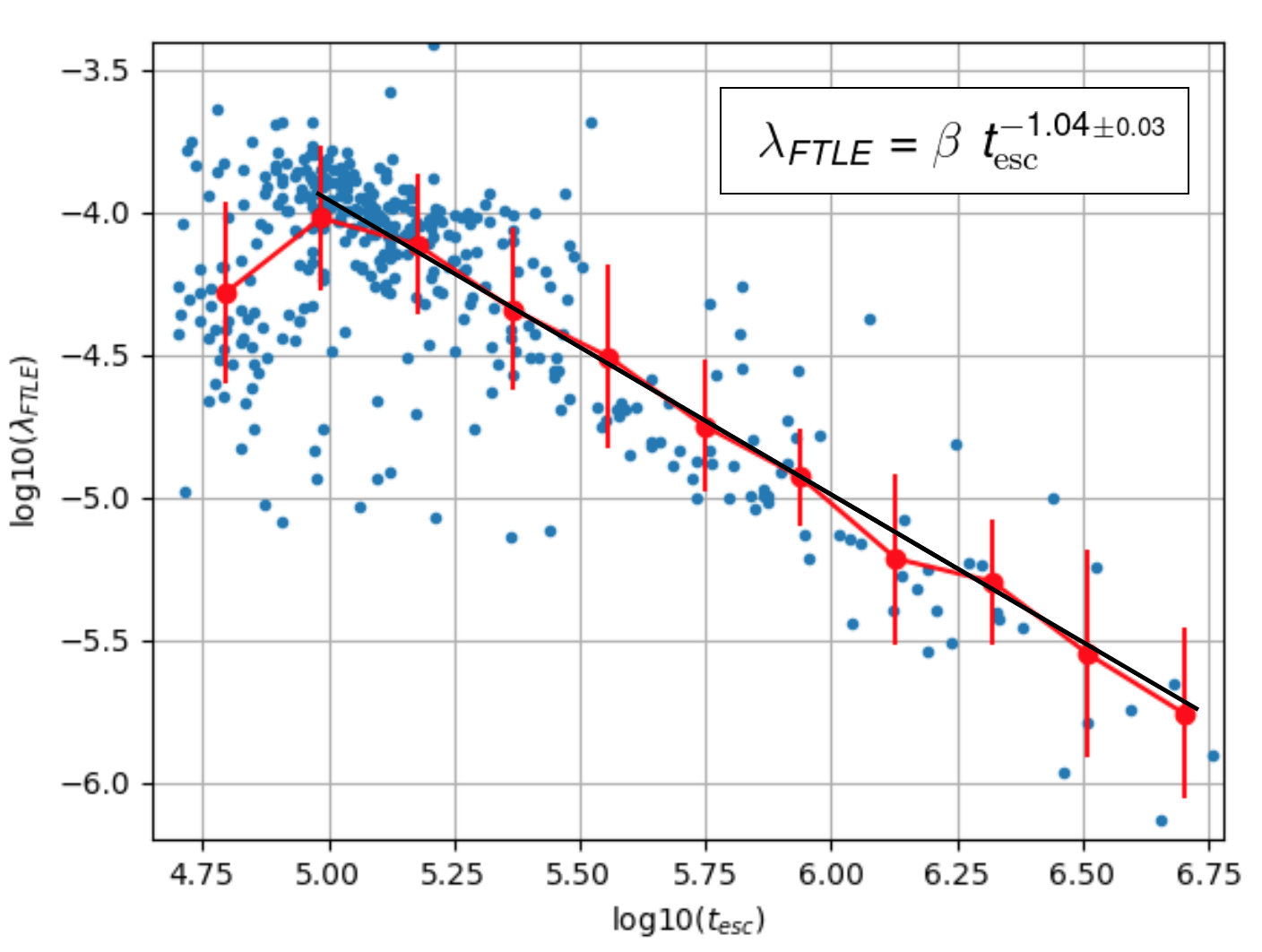}
	\caption{ \textit{Unperturbed system}: The finite-time Lyapunov exponent, $\lambda_{\tiny{FTLE}}$, vs. the escape time from the system, $t_{esc}$, for the unpertubed system. The blue points denote the specific values for each set of particles, which correspond to different initial conditions. The profile is denoted by the red points and the red line connecting them. The vertical red error bars correspond to one standard deviation. Note that from $t_{esc}\sim10^5$ to the maximum escape time, the distribution follows a power-law–like behavior. The fit for the power law of the profile is given by the Eq. \ref{eq:powerlaw}, which shows a power of -1.04.}
	\label{fig:finitelyapunovvsescapetimeunperturbedchi2500}
\end{figure*}

\begin{figure}
	\centering
	\includegraphics[width=1\linewidth]{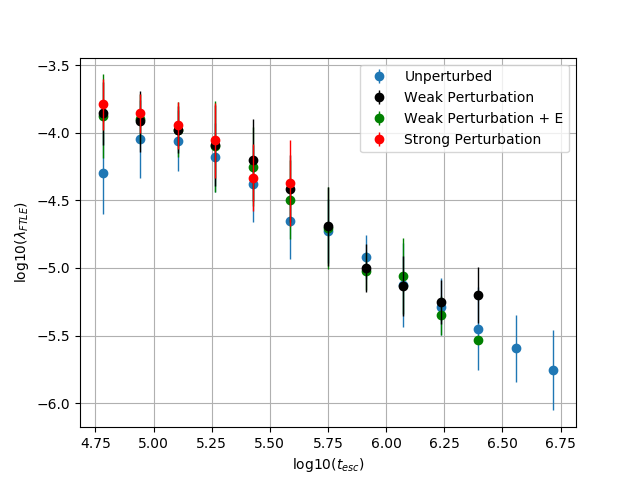}
	\caption{  \textit{Comparison of perturbed systems.} The finite-time Lyapunov exponent, $\lambda_{\tiny{FTLE}}$, vs. the escape time from the system, $t_{esc}$, for four different cases. The blue points represent the unperturbed system shown in Figure \ref{fig:finitelyapunovvsescapetimeunperturbedchi2500}. The black points correspond to the profile of the weak-perturbation system, the green ones show the weak perturbation plus electric field, and the red ones the strong-perturbation system. Section \ref{sec:mbottle} gives the description for each magnetic field configuration. Note that once perturbations are introduced in the system, the overall distribution of particles in the different categories changes; nonetheless, the power-law behavior and slope remain the same. }
	\label{fig:finitelyapunovvsescapetimefourcaseschi2500}
\end{figure}

\begin{figure}
	\centering
	\includegraphics[width=1\linewidth]{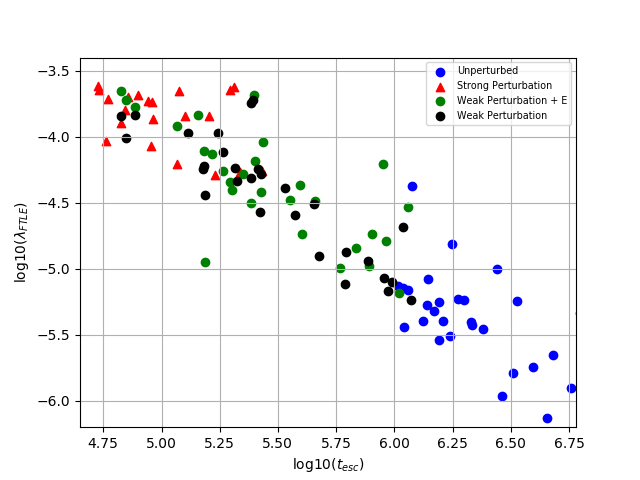}
	\caption{ \textit{Migration.} The finite-time Lyapunov exponent, $\lambda_{\tiny{FTLE}}$, vs. the escape time from the system, $t_{esc}$, for the same set of reference trajectories. The original set of particles in the unperturbed system is shown in blue. The black points show the same set of reference trajectories but subjected to a weak perturbation. The green points correspond to these particles in a weak perturbation plus electric field system, and the red points represent the particles affected by the strong perturbation. Note that even though the reference trajectories have the same initial conditions, these particles originally in the regular category in the perturbed system can reduce their escape times by a factor of two if affected by a strong perturbation. But in doing so, they still show the same power-law behavior. }
	\label{fig:finitelyapunovvsescapetimemigration}
\end{figure}

In Figure \ref{fig:finitelyapunovvsescapetimeunperturbedchi2500}, the data for each set of particles is divided into eleven bins, according to their escape times $t_{esc}$. Then, an average for each bin is calculated and denoted by a red point in the figure. Subsequently, the red points are connected by a red line to show the trend for the profile. This profiling is also done for Figures \ref{fig:finitelyapunovvsescapetimefourcaseschi2500} and \ref{fig:finitelyapunovvsescapetimemigration} for the perturbed cases. 


Given this profile, it is found in Figure \ref{fig:finitelyapunovvsescapetimeunperturbedchi2500} that the distribution exhibits a power-law behavior after reaching the maximum values for $\lambda_{\tiny{FTLE}}$ at 
$10^{-4.0}$. The power law extends from $t_{esc}\sim10^5$ to the maximum escape times for the system. 
The fit for the profile is given by the expression


\begin{equation}
\lambda_{\tiny{FTLE}} = \beta\, t_{esc}^{-1.04 \pm 0.03} .
\label{eq:powerlaw}
\end{equation}
This fit has an $R^2$ value of 0.995 and a scaling value $\beta = 10^{1.24 \pm 0.15}$.

One important feature to notice is that this slope is the same as that exhibited in the perturbed cases in Figures \ref{fig:finitelyapunovvsescapetimefourcaseschi2500} and \ref{fig:finitelyapunovvsescapetimemigration}. 
In Fig. \ref{fig:finitelyapunovvsescapetimefourcaseschi2500}, we can see that the data for the weak perturbation and the strong perturbation all show the same power-law behavior with a slope of approximately minus unity. 
This feature 
is even more clear in Figure~\ref{fig:finitelyapunovvsescapetimemigration}. The particles denoted in blue have the same initial conditions as the others but are subjected to different perturbations. Even though these perturbations affect the final escape time that they have, their behavior is still along the same power law. 

In Fig. \ref{fig:finitelyapunovvsescapetimefourcaseschi2500}, the different sets of particles are subjected to various magnetic field configurations, as described in Section \ref{sec:mbottle}. For each set of particles, the initial conditions for the reference trajectories are kept the same, so that any differences will arise from the various perturbations introduced in the system. 
From this figure, we can see that if these perturbations are present in the system, the distribution of particles in the different categories of the FTLE changes. However, the same power-law behavior remains. 
The most evident features that changed in this perturbed case are that there are no longer particles in the trapped category, and particles are rearranged along the power law depending on how strong the perturbation they experience is. 
For example, if a weak perturbation is introduced, we can still see that there are particles in the regular region. However, if a strong perturbation is present, particles tend to leave the system at a faster pace; therefore, the regular and trapped categories will be depleted of particles. 

This migration of particles from one category to another one is shown in Figure \ref{fig:finitelyapunovvsescapetimemigration}. Here particles originally in the regular category of the unperturbed system (denoted in blue in the figure) were subjected to the various perturbations. The reference particles' initial momentum and position are kept the same. It is shown here that these sets of particles escape the system more quickly, moving to the irregular and intermediate categories when a weak perturbation is present. In the presence of the strong perturbation, almost all of them move to the intermediate category. 
Additionally, even though the particles change their categories and move to shorter escape times, they do so in a manner that still complies with the power-law behavior.

\begin{figure}
	\centering
	\includegraphics[width=1\linewidth]{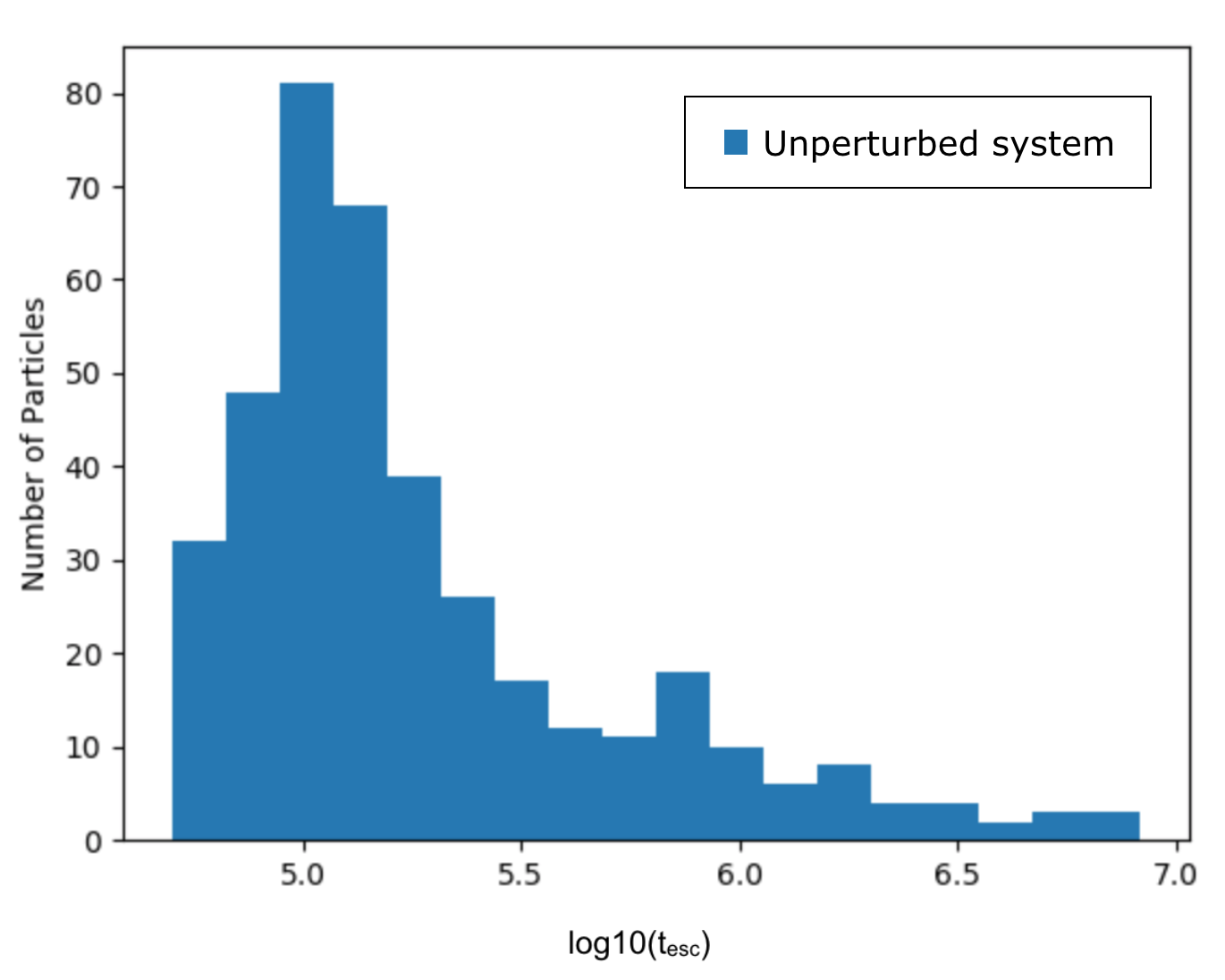}
	\caption{ \textit{Histogram for final escape times.} Number of sets of particles vs. the escape time from the system, $t_{esc}$, in the unpertubed system.  Note that if we interpret this plot as a probability distribution, particles are more likely to have escape times around $t_{esc} = 10^5$, which is consistent with the migration depicted in Figure \ref{fig:finitelyapunovvsescapetimemigration}.}
	\label{fig:histoselectedfinaltimesloglmandtrfunpert}
\end{figure}

The histogram depicted in Figure \ref{fig:histoselectedfinaltimesloglmandtrfunpert} shows the distribution of particles for the final escape times in the unperturbed system. 
This histogram can be interpreted as a probability distribution plot for escape times, where the most likely scenario is around  $t_{esc} = 10^5$. This most probable case is consistent with the migration plot in Figure \ref{fig:finitelyapunovvsescapetimemigration}. In that figure, when perturbations are introduced, the particles tend to move to shorter times and accumulate around the  $t_{esc} = 10^5$ range for the most extreme case. Therefore, a histogram such as the one in Figure \ref{fig:histoselectedfinaltimesloglmandtrfunpert} could be used as a predictor for the expected behavior of a set of particles when the system is perturbed. For example, a particle that has a long escape time will have a tendency to move to a more likely scenario when the perturbation is introduced.

\subsection{Maps}
\label{ssec:arrivalmaps}

\begin{figure*}
	\centering
	\includegraphics[width=1\linewidth]{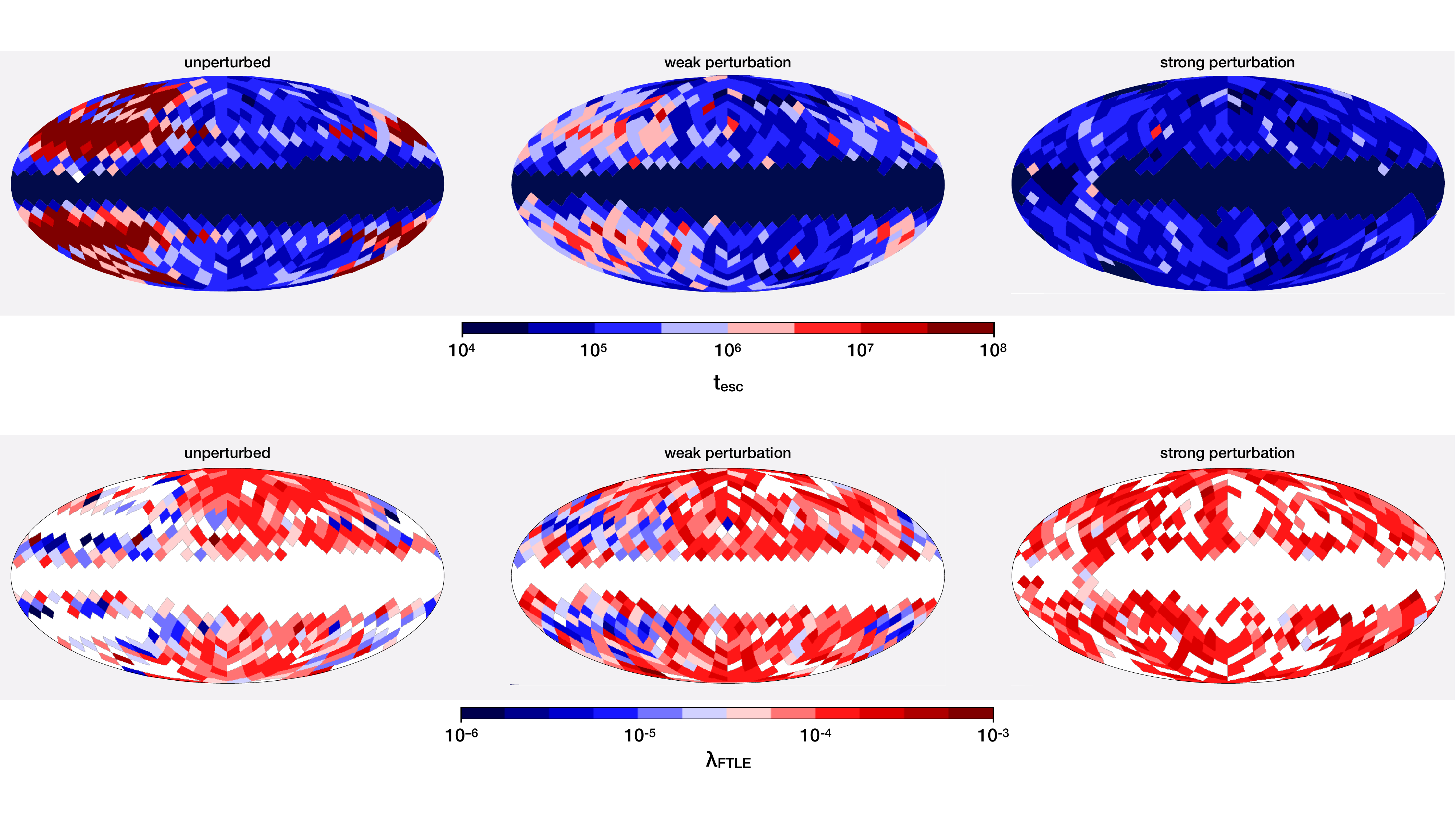}
	\caption{ \textit{Maps.} The location of each pixel in a map corresponds to the arrival direction of a reference particle, and the values for each pixel represent the escape time or finite-time Lyapunov exponent. The top panel corresponds to the escape times for the unperturbed, weakly perturbed, and strongly perturbed systems, respectively. The bottom panel corresponds to the finite-time Lyapunov exponent, $\lambda_{\tiny{FTLE}}$, for those systems. The white pixels in the bottom panel are for particles that are not chaotic. These maps correspond to a visual representation of the different chaotic behaviors and how they are distributed spatially. For instance, we can see areas of the unperturbed map where particles are more chaotic (denoted in redder colors in the bottom panel) and areas in the vicinity of the stability region that are less chaotic (darker blue). As the time-perturbation gets stronger, we can see the progression of the maps, where the more chaotic particles start to populate larger regions of the map. In the case of the heliospheric effects, we can expect that the maps would look like the ones in the middle panel. With a weak perturbation, there is a mix of how chaotic particles can be, and there will be regions where you can have more significant variations due to the chaotic nature of the particles in it.}
	\label{fig:figure-skymaps}
\end{figure*}

\begin{figure*}
	\centering
	\includegraphics[width=1\linewidth]{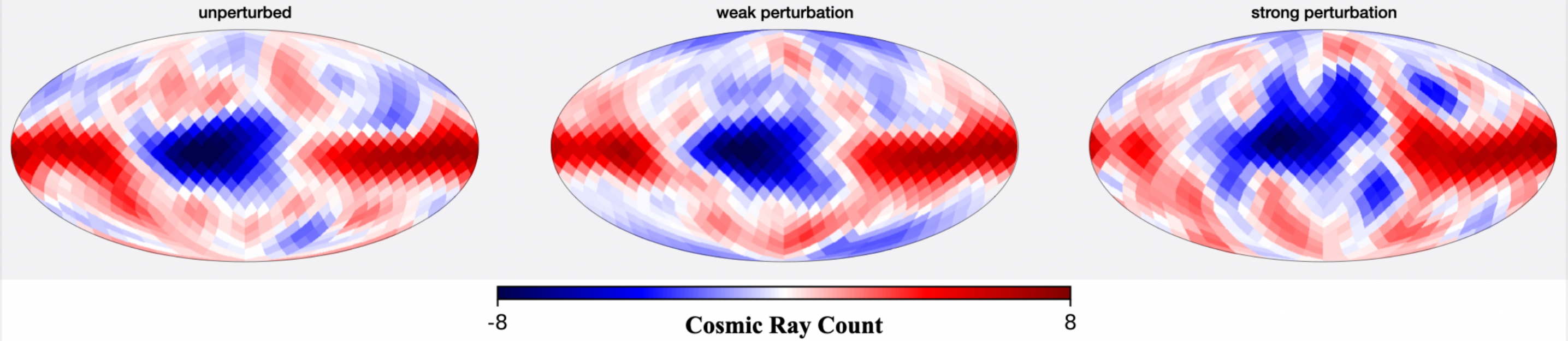}
	\caption{ \textit{Sky Maps of Arrival Direction Distribution.} Maps of cosmic ray arrival direction distribution of 1 TeV protons for the unperturbed, weakly perturbed, and strongly perturbed systems, respectively. The values for each pixel correspond to cosmic ray count where each particle event is multiplied by its dipole weight as explained in Section~\ref{ssec:arrivalmaps}. Gaussian smoothing with $\sigma=0.15$ rad was used. The weak perturbation system approximately represents the large-scale magnetic field conditions in the heliosphere and the variability of solar wind properties along the heliosphere beyond the termination shock.
}
	\label{fig:figure-directionmaps}
\end{figure*}

Following the calculation of the finite-time Lyapunov exponents and escape times for the sets of trajectories, we proceed to plot them in an arrival distribution map (see Fig. \ref{fig:figure-skymaps}). The location of each pixel corresponds to the arrival direction of a
reference particle, and the value assigned to each pixel indicates the FTLE (Fig. \ref{fig:figure-skymaps} top panel), or the escape time
(Fig. \ref{fig:figure-skymaps} bottom panel).  Therefore, the maps correspond to a visual representation of the different chaotic behaviors and how they are distributed.
In these arrival distribution maps, we can identify that there are regions of stability where the particles are trapped within the maximum integration time (denoted in deep red in the top panel of Fig.~\ref{fig:figure-skymaps}). Originating from those regions, gradients from longer to shorter escape times
appear. Since $t_{esc}$ is related to the FTLE, the chaotic behavior of particles follows this trend as well; see Fig. \ref{fig:figure-skymaps} bottom panel.

Particles in the power law are located outside that stability region and populate the rest of the areas of the map (except for the region around the equator). Regular particles with very long trajectories are located at the margin of the stability region, with moderate Lyapunov exponents and manifesting a metastable behavior.  Then, we can see a transition to the higher Lyapunov exponent for particles with shorter times (intermediate and irregular particles), which are more spatially distributed in the map. 

The particles in the equator remain in the system for a particularly short time and, therefore, do not contribute to the Lyapunov maps. Their exponent is taken to be zero under these conditions. 

For the case of the maps with perturbations present (see columns 2 and 3 in Fig.~\ref{fig:figure-skymaps}), we see significant changes that start to appear as the perturbation goes from weak to strong. The most obvious change is that there is not a stability region in these cases, but still the transitions from lower FTLE to higher values are present. For the weak perturbation, there is still a variety of behaviors and values for the FTLE, yet, for the strong perturbation, we see that it is more restrictive on the values that the exponents can take, and it is basically populated by one type of particles, as we have seen in the $\lambda_{\tiny{FTLE}}$, vs. the escape time  $t_{esc}$ plots.  The FTLE values for these particles are uniformly distributed in the map as well. 

The sky maps of arrival direction distribution are plotted in Figure \ref{fig:figure-directionmaps}. As delineated in Section~\ref{sec:propChaos}, generating the sky maps of the particles' arrival distribution involves integrating their trajectories back in time. The reference particles are assumed to uniformly originate from a single point, which is considered to be Earth. For each reference particle, we randomly inject a set of 10 particles on the surface of a sphere with a radius of 0.01 and identical momenta (see Figure \ref{fig:method}). These particles travel outward until they reach the maximum radius, a sphere with a radius of $r_{max}$ = 12500 in code units, corresponding to 2500 AU. Each trajectory's position and momentum direction are recorded in this outer sphere. Subsequently, the trajectories are inverted in time from their positions on the outer sphere to their origin, i.e., Earth.\footnote{See, for example, \cite{barquero_2016} on how this back-propagating process is done in the context of turbulence in the ISM.}

While inverting time on the numerically calculated trajectories, we introduce a dipole gradient distribution as a weight on each trajectory at the crossing point on the outer sphere. This weight is calculated based on the angle of the particle direction at the crossing point in relation to the dipole gradient, $\vec{p} \cdot \vec{B_d} = pB_d cos\theta$. We then determine the arrival distribution of these forward-propagated trajectories at Earth. In this work, the dipole direction is chosen in the (1,0,0) direction. Since this is a toy model, the dipole direction can be freely chosen. In a real scenario, this dipole can be aligned with the direction of the mean magnetic field in the ISM.

The most prominent area in the arrival direction sky maps is the equator, but the most significant changes across the maps are located outside this region. 
When the three maps for the various perturbation levels are compared, the progression in the morphology of the anisotropy is clear. 
These perturbations do not entirely alter the large-scale morphology; instead, the changes are localized, resulting in the emergence or accentuation of specific features. The most noticeable change occurs when the strong perturbation is introduced, where the features present in the two maps on the left are given more angular power. Such changes can be expected for perturbations of this nature as the large-scale characteristics of the system configuration are predominantly preserved
(see Section~\ref{ssec:timeperturb} for a discussion on the aspects that contribute to the perturbations' influence, and Section~\ref{sec:observations} for a description of the connection with the anisotropy observations).


\section{Discussion}
\label{sec:disc}


A new method to calculate the chaotic behavior of particles' trajectories in bound systems has been developed. This method is based on the calculation of the finite-time Lyapunov exponent, a quantity that is adaptable to bounded conditions; see Section \ref{sec:chaos}.
The FTLE is used to characterize the particles' behavior while bounded in the system but also to capture changes in behavior and transitional states. We can envision an example where particles inside a coherent magnetic structure experience an exponential divergence, then move to a steady bounded state, and later escape. After this escape, they can propagate in a larger field and then encounter another structure where they could get bounded again. 
Consequently, instead of tracking just one regime, as would be the case for calculating a specific diffusive state, the FTLE can adapt to the changing conditions. Therefore, using it in conjunction with a well-defined diffusive state can yield a more comprehensive understanding of the CR propagation.

To describe the particles' behavior in bound systems, we constructed a toy model as our propagating medium. It consists of a magnetic bottle with time-perturbations added (see Section \ref{sec:mbottle} for details). The specific parameters used in this work are based on the heliospheric magnetic configuration; nonetheless, it can be adjusted to fit different structures, as it will be discussed in Section \ref{sec:outlook}. This model captures the heliospheric large-scale magnetic features. One of these features is the magnetic mirroring effect between its flanks due to the draping of the interstellar magnetic field. The other effect is due to the solar cycles and its magnetic field variation in polarity.

\subsection{Chaotic Behavior}

As particles propagate in this system, they display chaotic behavior, as shown in Section \ref{sec:results}. 
We found that the degree of chaos of the trajectories is correlated to the particles' escape time from the system. This relation is given by a power law (see Section \ref{ssec:ftle}), and depending on that, different behavioral regimes exist. 


Particles can undergo four phases within the escape time that determine their behavioral category. However, a particle does not necessarily experience all of these phases. The initial phase is when divergence has not happened yet. Overall shorter trajectories experience almost no separation for just a brief interval compared to longer trajectories, for which this period is considerably more prolonged, as seen in Figure \ref{fig:sepratiohisto206216}. The next phase is the divergence stage, where longer trajectories diverge at a slower rate, hence their low FTLE values.  
A penultimate phase, observed in long trajectories, corresponds to an extended slow-divergence period and an approximately constant separation behavior. The particles remain bounded in the system for a long time. The final phase is the escape, where particles leave the confined system. 

Different CR behavioral categories can be identified in these bound systems, as discussed in Section \ref{sec:results}. These regimes depend on multiple elements, such as the trajectories' specific characteristics, the phases particles experienced, and their corresponding chaotic behavior. 

The first of these categories is for particles that leave the system in a short time, and therefore, they cannot develop any chaos. These transient particles have very smooth trajectories and do not experience significant variations in the magnetic field. In a realistic environment,  these particles will most likely trace out the magnetic field outside the bound system. For example, particles that enter the heliosphere through the nose, in general, spend a very short time in the heliosphere before being detected at Earth. Therefore, they could experience the least deviations and be more closely connected to their original direction in the interstellar medium. 



The second type of particle includes those in an intermediate state. These particles do not achieve their maximum chaotic potential since they escape the system before doing so. Their divergent behavior is very explosive and occurs in a short period. They encounter regions where the magnetic fields vary vastly, and as a consequence, their escape times are shortened.
Therefore, these particles are not likely to be able to trace out the magnetic field outside the magnetic structure. Their behavior is in between the transient and the power-law regime in the system.

The next category, and where the vast majority of our particles reside, is the power-law regime. When we explore the relation between the FTLEs and the escape times, a power law emerges; see Fig.~\ref{fig:finitelyapunovvsescapetimeunperturbedchi2500}. 
In this power law, the Lyapunov exponent follows the inverse of the escape time; see Eq. \ref{eq:powerlaw}.
As we have seen in Figure~\ref{fig:sepratiohisto206216}, particles that diverge too quickly are unable to maintain that rapid divergence for an extended period, due to the finite size of the system, and escape rapidly. On the other hand, a slower divergence gives particles plenty of time to spend bounded in the system before leaving it. 
Therefore, shorter trajectories have higher FTLEs, and longer trajectories display less chaotic behavior.


Different factors contribute to the power-law relation of the FTLE vs. escape time. These elements are based on the connection between the magnetic field geometry and the test particle's energy. 

One of the elements that conditions the particles' behavior in the system is the ratio $\omega_{bounce}/\omega_g$, where $\omega_{bounce}$
is the bouncing frequency between the mirrors, and $\omega_g$ is the gyro-frequency. 
This ratio's value is smaller for trapped trajectories than those from other categories of particles, and it does not vary much in general. This ratio has a more significant variation and a higher value for the more chaotic particles.

There are also limitations set on the maximum FTLE given by the configuration of the system. In Figure~\ref{fig:finitelyapunovvsescapetimeunperturbedchi2500}, it is shown that the highest values for the FTLE have magnitudes of $10^{-4}$ and are found in the vicinities of $t_{esc}=10^5$. The maximum value achievable for the FTLE for which the particles can still be bounded is $4.90 \times 10^{-4}$. It is a constraint given by the system's physical dimensions. Its value is determined by the maximum separation that particles can achieve while still trapped and the time that it takes to achieve it, as denoted in Eq. \ref{FTLE}. For instance, a particle with a $ \lambda_{\tiny{FTLE}} =10^{-3} $ would not be able to stay in the bounded system since its divergence is so extreme. 

Therefore, for a real finite system, one could know its chaotic potential just by the system's overall dimensions compared to the characteristics of the impinging particles. This indication could be beneficial if there is a collection of similar coherent structures. Each of them can contribute to the general behavior, and we could predict their effects in the overall propagation. For example, we would expect these conditions in certain interstellar medium regions, so that we could see how the diffusion of CRs is affected by the presence of a collection of coherent magnetic structures.

Another element present in the FTLE vs. escape time plots is an inflection point in the profile. For the unperturbed system, an inflection point is found in the profile (Fig. \ref{fig:finitelyapunovvsescapetimeunperturbedchi2500}) where the maximum values for the FTLE occur ($t_{esc}=10^5$).  This inflection point at the profile's maximum values is expected since the intermediate particles displayed a very explosive divergence and are not stable enough to remain in the system. The particles at the inflection position are the most chaotic that were able to thoroughly diverge in the system. From that point on, particles will start to diverge at a slower pace. However, the exact location of the inflection is dependent on the perturbations that act on the system. For the perturbed cases, this inflection point is located at a shorter escape time and a slightly higher value of the $\lambda_{\tiny{FTLE}}$, as shown in Figure~\ref{fig:finitelyapunovvsescapetimefourcaseschi2500}. Since the perturbations are a source of chaos in the trajectories, this decrease in escape time and increase in the FTLE values of the inflection point is expected. Nonetheless, this shift in the inflection point location is restricted since its maximum value is already determined by the system's dimensions, as mentioned before.


\subsection{Effects of Time Perturbations}
\label{ssec:timeperturb}

An important attribute that this system presents is that if the system is perturbed, the particle's behavior still falls along the same power law. 
As shown in Figure \ref{fig:finitelyapunovvsescapetimefourcaseschi2500} and more clearly in Figure \ref{fig:finitelyapunovvsescapetimemigration}, once a perturbation affects a set of particles, the particles' behavior becomes more chaotic, but it follows the same power law as in the unperturbed case. Multiple factors can contribute to this phenomenon, from the perturbation's spatial dimensions to the overall magnitude of the perturbations' magnetic field compared to those from the magnetic bottle. 

One of the aspects that could contribute to the permanence of the particles in the power law is that the time-perturbation is not very extended spatially compared to the bottle's dimensions. The particles are essentially still on the same system configuration, and the perturbation will solely drive them to another possible path of the same system. Accordingly, they will follow the same power law. Moreover, as we have seen in Figure \ref{fig:histoselectedfinaltimesloglmandtrfunpert}, the region around $t_{esc} = 10^{5}$ is the most likely scenario; consequently, as the system is perturbed, it will be driven to this most probable case. 


Another contributing factor is that the perturbation is not strong enough to change the whole behavior. Therefore, the first order in terms of the system is the magnetic bottle. Suppose the magnitude of perturbation was higher, or perhaps its extension was ampler. In that case, the bottle's magnetic field will be secondary, and the overall behavior, including the slope in FTLE vs. escape time, will change. This point is also related to the fact that the perturbation that the particles experience does not vary that much as the particles' trajectory progresses. For example, for the strong perturbation and a particle with a final time of $10^5$, the perturbation only moves 6.6 AU before the particle escapes the system. 

Nonetheless, even if the perturbation does not deviate the particles from this power law, the cumulative effect of the magnetic bottle plus the perturbation does create severe chaotic changes. This idea points to the fact that even small changes can have significant effects. 

Given the invariance under these perturbations that the Lyapunov-exponent-escape-time relation displays, this exact power law could prove to be an intrinsic property of the system. Therefore, a more profound question arises: could such a power law be present for other similarly arranged magnetic configurations? Examining the robustness of the power-law distribution across various magnetic configurations with a range of perturbations is an essential area for future research. By methodically exploring the impact of different magnetic field configurations and perturbations on the correlation between FTLE and escape time, we can better comprehend this fundamental dynamics on cosmic ray propagation. This understanding can provide insights into the spatial distributions and their implications.





\section{Connection with the observations}
\label{sec:observations}

Another crucial aspect to consider is how chaos and magnetically connected areas in the system can affect the observations. One of the most important consequences that chaos and the particles' interaction with the heliospheric system can have is that it can potentially create time-variability on the cosmic ray anisotropy maps.

Chaos' canonical idea is that even a small change can have tremendous consequences. As we have shown in our analysis, cosmic ray trajectories are vastly affected by our heliosphere, and they are, in their majority, chaotic. Therefore, a slight change in the particles' initial conditions or the magnetic field through which they traverse can produce significant differences. Specifically, here, the most evident observable changes are temporal variations that can occur in the arrival spatial distribution. These changes are physically associated with the heliosphere's topological features, dimensions, and 11-year solar cycles. Consequently, we anticipate these variations to become discernible within the timeframe of the solar cycle. In terms of scales and energies, particles with rigidities ranging from 1 to 10 TV are susceptible to the influence of the mirror configuration created by the interstellar magnetic field lines draping around the flanks of the heliosphere. This energy range aligns with the combined observations of HAWC and IceCube, which cover the full sky at a median energy of 10 TeV~\citep{Abeysekara_2019}, providing the basis for our study.

When assessing the effects on the CR arrival direction's anisotropy, it is essential to examine the contributions of the magnetic field variations and discern the distinct roles played by each category of particles. For example, if the perturbation on the magnetic structure is strong, the original distribution can be completely rearranged. Accordingly, the distribution will lose memory of its original form after one Lyapunov time. For a more moderate perturbation, as the one analyzed here or in the case of the heliosphere, the effect is significant, yet some of the original distribution remains. Therefore, realistically, we could expect three situations: (1) the map's overall spatial distribution is shifted slightly, (2) parts of it are shifted, or (3) regions of the maps are changed while others are practically unaffected. Our chaotic and arrival direction map distributions (Figures \ref{fig:figure-skymaps} and \ref{fig:figure-directionmaps}) point to the latter cases as the closest to what the heliosphere can cause since a strong perturbation is needed for the former case, where the whole map distribution completely changes.

Therefore, in the presence of temporal variability within the maps, the primary contributors to this variation, and consequently the most significantly impacted particles, are those exhibiting chaotic behavior. Specifically, this pertains to particles in the intermediate and power-law categories, as expounded upon in subsection \ref{ssec:class}. Notably, transient particles may exhibit a distinct response, potentially remaining unaffected or demonstrating a collective motion characterized by a gradual drift. This phenomenon stems from their brief time within the system, limiting the opportunity to display chaotic trajectories. Moreover, the trajectories of transient particles exhibit robust magnetic connections to the external environment, establishing a direct and expeditious link to outside the system (see Figure \ref{fig:figure-trajectories}, top left panel, for an example trajectory). As a result, transient trajectories could directly map the initial and final phase space configuration of the area they populate. Chaotic trajectories, on the other hand, have a more complex scenario. For instance, considering the central bottom panel map of Figure \ref{fig:figure-skymaps} as our point of reference, we can divide it into three regions. The white middle region accommodates the aforementioned transient particles, the blue regions contain particles exhibiting mild chaos, and the red regions encompass highly chaotic particles. Due to the complexity of the interaction between magnetic fields and particles, no sharp boundaries separate each region. As a result, gradients can be observed when transitioning from one region to another. Therefore, if similar sectorial morphology were present in the real observations, we would expect areas in the sky (akin to the ones in red in Figure \ref{fig:figure-skymaps}) to exhibit more significant variations compared to their surroundings.

Preliminary results from 11 years of IceCube data (Figure 6 in \cite{McNallyICRC2023}) show a possible time variability in their anisotropy observations, where the temporal variations change depending on the region of the sky. Nevertheless, a more comprehensive investigation is imperative to determine the statistical significance of these observed effects. One important aspect of these observations will be determining whether the effect is seen on the smaller angular scales or if the large scales are impacted. It is crucial to underscore that the data collection period encompasses an entire solar cycle; thus, we expect the influence of the heliosphere to be present, adding a vital dimension to the analysis. The heliosphere has an intrinsic directionality that affects cosmic rays differently depending on where they enter it. In \cite{barquero_2017}, particles exhibit a greater spatial dispersion in the interstellar-wind downstream direction due to the elongated heliospheric tail, in contrast to their distribution in the upstream direction. This feature results in preferential directions from which particles tend to be more chaotic than others. This eventually translates into changes in the arrival maps, which are not uniformly distributed. Therefore, sectors in the map change separately from others, creating a time variation that could be detected. Consequently, this result points to the idea that time-variability in the maps is essential to understanding the CR anisotropy's overall processes and magnetic structures involved.

In this work, we employed a toy model of the heliosphere, which has provided a deeper look at the consequences that mirroring and temporal perturbations can cause on particle trajectories and spatial distribution. Additionally, we developed a method to quantitatively assess the degree of chaos in cosmic ray particle trajectories. However, to establish direct correlations with observable features, employing a more realistic representation of the heliosphere becomes necessary. Hence, the primary objective of an ongoing study and future publication is to apply the methodologies, theoretical framework, and findings elucidated in this work while incorporating an MHD-kinetic model of the heliosphere \citep{Pogorelov_2015} to examine the effects on anisotropy maps.


\section{Outlook}
\label{sec:outlook}

As cosmic rays propagate, they encounter magnetic structures that could trap them temporarily and induce chaotic behavior in their trajectories. The model that we develop in this study can be used to represent a variety of magnetic structures and magnetic processes that lead to such interactions. In addition, the method we construct here to characterize chaotic behavior based on the FTLE can be applied to these cosmic-ray trapping scenarios. However, it can also be used when particles move rapidly from one environment to another, and when their trajectories experience a change of first-order effects on them.

The rate at which CRs have such bounding interactions depends on several factors. One of the determinant elements is the particle's energy or, more specifically, the particle's rigidity (defined as R = E/Ze, where E is the energy, Z the atomic number, and e the electric charge). Depending on the particle's rigidity and its corresponding gyroradius in a specific magnetic field, it can experience strong effects from magnetic structures of a similar scale. If the gyroradius is smaller than the coherent structure, it will be affected by the magnetic field's overall geometry. If it is larger than the magnetic perturbation, accumulating effects will be felt.

The trapping of cosmic rays with their consequent chaotic effects can happen at various scales, including those of the interstellar medium and the intercluster medium~\citep{farrar_2019,jansson_2012,tharakkal_2022,kulsrud_1975}. Different types of processes can also be behind it. Therefore, from its place of origin to its detection on Earth, a particle can be affected by multiple coherent magnetic structures. Nonetheless, we can expect the number of interactions to vary for a distribution of particles at different rigidities or injected in different places in the galaxy.

For instance, we can visualize two distinct extreme scenarios for particles at different rigidities. For a PeV proton injected into the galaxy, the motion could be dominated by the trapping of a few specific magnetic structures at the time of the detection. However, we could consider, for example, a TeV proton, for which the situation may look completely different. These cosmic rays with lower energy would have more chances to encounter structures that can affect them. We could also conceive that differences will arise if particles are injected in a relatively quiet place in the galaxy instead of a very turbulent and energetic site. These scenarios can profoundly affect the arrival distribution on Earth, since our surroundings could select particles with specific rigidities.

We can expect these interactions to arise from the cosmic rays' interplay with very well-defined structures such as the heliosphere or as a result of more basic phenomena, e.g., structures that appear due to spatial intermittency. 

The magnetic configuration that we use as inspiration for our toy model system is the heliosphere. Here we have a significant source of mirroring effects between the flanks of the heliosphere. Moreover, there are a variety of perturbation sources. One of them is the one that comes as a result of the solar cycles that we described in this study. The particles' chaotic behavior could change depending on the phase in the solar cycle in which the system is. The perturbation phase may distribute chaotic and nonchaotic CRs differently as a function of the phase of magnetic perturbation. This effect is because the surrounding space has different magnetic field polarities. Therefore, a time variability could come from the existence of a perturbation, but the definite characteristics of it can have a significant impact, in this case, the polarity.

Other perturbation sources are the instabilities at the interface between the ISM and the heliosphere, turbulence, and the motion that the heliospheric flanks have relative to the inner heliosphere. This latter motion is described as having a speed of 10-100 AU per year~\citep{1996JGR...10121639Z,2021ApJ...922..181O}, which could be a consistent source of variability and directionality in the maps.

Similarly, we could expect mirroring effects or trapping in more extensive structures, such as the Local Interstellar Cloud (LIC) or the Local Bubble (or Local Cavity) (see ~\cite{frisch_1998, abt_2015, Nojiri_Globus_2024} for details). The LIC has an extension of 30 light-years and the Local Bubble of 300 light-years. So their influence can span over particles with rigidities in the $10^{17}$ V region. 

As mentioned before, spatial intermittency plays an essential role when dealing with the creation of coherent structures~\citep{Matthaeus_2015}. These structures affect the cosmic-ray propagation and a possible diffusive state. These effects can be involved in multiple scenarios at different scales, such as the interstellar medium, solar processes, and magnetospheres, as well as fundamental physics since empirically their origin resides in the nonlinear dynamics of turbulence.

\section{Summary \& Conclusions}
\label{sec:conclusions}

In this work, we have explored the possibility that chaotic behavior can originate from the interaction between cosmic rays and magnetic structures that could vary in time. These structures can temporarily trap cosmic rays in them, leading to their chaotic behavior. We also present the potential consequences that it can have on the cosmic ray arrival observations (Section \ref{sec:observations}).

The toy model developed in this work, consisting of a magnetic bottle with time perturbations (Section \ref{sec:mbottle}), reproduces the large-scale magnetic features of the heliosphere. One of these features is the magnetic mirroring effect between its flanks, resulting from the draping of the interstellar magnetic field. Furthermore, the model also encompasses the variations in solar cycles and their consequential changes in magnetic field polarity, adding another dimension to its representation. While we have applied the toy model to the heliosphere, there is a wide range of potential magnetic configurations where this model will be relevant, including spatial intermittency and mirroring in more extensive structures such as the Local Bubble (Section \ref{sec:outlook}).

The magnetic bottle and the perturbations cause significant changes to the trajectories of the cosmic rays. To fully describe these changes, we developed a new method for characterizing chaotic trajectories based on the finite-time Lyapunov exponent, FTLE. This quantity is especially advantageous since it can adapt to transitory behavior, including the temporary trapping conditions in this model as explained in Sections \ref{sec:lya}, \ref{sec:disc}, and \ref{sec:outlook}.


The results are summarized as follows:	

\begin{itemize}
	
\item As particles propagate in the system, they display chaotic behavior. Our results show that the finite-time Lyapunov exponent, a quantity that indicates the chaotic behavior of a trajectory, is related to the escape time of the system. This relation is given by a power law, $\lambda_{\tiny{FTLE}} \sim t_{esc}^{-1.04}$. Thus, it indicates that particles exhibiting shorter escape times display higher levels of chaos compared to cosmic rays that remain within the system for longer durations. Therefore, we can classify the particle's trajectories depending on their degree of chaos.

\item The maps of arrival distribution display areas where the chaotic features vary significantly; these changes can be the basis for time variability in the maps. The particles that will mainly drive this variation and be the most affected are those in the intermediate and power-law categories. These are the most chaotic ones, with high Lyapunov exponents.  In our arrival maps, chaotic trajectories are divided into two regions, one containing particles exhibiting mild chaos and another encompassing highly chaotic particles. Given the intricate interplay between magnetic fields and particles, there are no clear-cut boundaries between these regions. As a result, gradients can be observed when transitioning from one region to another. Consequently, if analogous sectorial morphology were observed in the real data, we would anticipate significant variations in certain sky regions compared to their surroundings. On the other hand, transient particles, i.e., particles that spend a very short time in the system, may not be affected or move as a whole, e.g., showing variation as a slow drift. Consequently, transient trajectories could directly map the ISM distribution and the observed configuration.

\item If time perturbations are introduced in the system, the particles' behavior becomes more chaotic, but it follows the same power law as in the unperturbed case. This power law could prove to be an intrinsic characteristic of the system; therefore, it can potentially offer further insights into the particle propagation dynamics beyond a simple diffusion scenario. One important aspect is that the time perturbations here are subtle and not very spatially extended, so they can influence the particles' behavior without completely changing the system's characteristics. This emphasizes the fact that even minor changes can have a significant impact. Testing the robustness of the power-law distribution in different magnetic configurations with a range of perturbations is a topic for future work.

\item Time variability could be an essential aspect of the observed CR anisotropy. In this work, the temporal variations affecting the spatial arrival distribution are physically attributed to the heliosphere's topological features, dimensions, and 11-year solar cycles. As these variations are linked to these heliospheric factors, we anticipate their discernibility within the timeframe of a solar cycle. Preliminary results from 11 years of IceCube data suggest potential time variability in anisotropy observations, varying across different sky regions. Further investigation is needed to ascertain these variations' statistical significance and scale dependence.

\item Furthermore, given the interplay of scales and rigidities, in the case of particles in the 1-10 TV range, the heliosphere represents the last dynamically relevant interaction before its detection. Therefore, we expect its effects to be more readily observable, especially since the combined full-sky observation by HAWC and IceCube is at a median energy of 10 TeV. 

\item Achieving direct correlations with observable features necessitates a more realistic representation of the heliosphere. Given the success of our FTLE techniques in the toy model, future work will utilize the methodologies and insights gained from this study on cosmic ray anisotropy maps from a more detailed MHD-kinetic model of the heliosphere. Consequently, it will enable us to thoroughly investigate the impacts of the heliospheric dynamics on the anisotropy maps.
	
\end{itemize}

\begin{acknowledgements}
VLB thanks the European Research Council (ERC) for support under the European Union's Horizon 2020 research and innovation program (grant No. 834203) and the Department of Astronomy at the University of Maryland, College Park.
\end{acknowledgements}

\end{document}